\documentclass[journal]{IEEEtran}

\usepackage{cite}
\ifCLASSINFOpdf
\usepackage[pdftex]{graphicx}

\usepackage{algorithm}
\usepackage{algcompatible}
\usepackage{algorithmicx}
\usepackage{amsmath}
\usepackage{amssymb}
\usepackage{booktabs,colortbl,multirow,multicol,ctable}
\usepackage{titlesec}
\titlespacing\section{0pt}{12pt plus 2pt minus 2pt}{5pt plus 2pt minus 2pt}
\titlespacing\subsection{0pt}{8pt plus 2pt minus 2pt}{4pt plus 2pt minus 2pt}
\usepackage{array}
\usepackage{caption}
\usepackage{subcaption}

\usepackage{color}
\usepackage{xcolor,colortbl}
\usepackage{soul}
\usepackage{cuted}
\makeatletter
\newcommand{\distas}[1]{\mathbin{\overset{#1}{\kern\z@\sim}}}
\newsavebox{\mybox}\newsavebox{\mysim}
\newcommand{\distras}[1]{
  \savebox{\mybox}{\hbox{\kern3pt$\scriptstyle#1$\kern3pt}}
 \savebox{\mysim}{\hbox{$\sim$}}
  \mathbin{\overset{#1}{\kern\z@\resizebox{\wd\mybox}{\ht\mysim}{$\sim$}}}
}
\renewcommand{\footnoterule}{%
 \vskip-3pt \hrule width 2in height 0.6pt depth \z@ \vskip1.6pt\relax
}
\makeatother
\newcommand{\E}{\mathbb{E}}
\newcommand{\R}{\mathbb{R}}
\newcommand{\x}{\mathbf{x}} 
 
\newcommand{\e}{\mathbf{e}}
\renewcommand{\r}{\mathbf{r}} 
\newcommand{\y}{\mathbf{y}} 
\newcommand{\z}{\mathbf{z}} 
\renewcommand{\u}{\mathbf{u}}

\definecolor{Gray}{gray}{0.85}
\newcolumntype{a}{>{\columncolor{Gray}}c}
\usepackage{tabularx}
\newcommand\tstrut{\rule{0pt}{2.4ex}}
\newcommand\bstrut{\rule[-1.0ex]{0pt}{0pt}}
\graphicspath{{fig/}}
\usepackage[finalnew]{trackchanges} 

\begin{document}
\setlength{\abovedisplayskip}{3pt}
\setlength{\belowdisplayskip}{2pt}
\setlength{\textfloatsep}{2pt}
\addtolength{\parskip}{-0.4mm}

\title{An Online Approach to Cyberattack Detection and Localization in Smart Grid}
\author{Dan~Li,~\IEEEmembership{Student~Member,~IEEE,}
        Nagi~Gebraeel,~%
        Kamran~Paynabar~%
        and~A.~P.~Sakis~Meliopoulos%
\thanks{Dan Li, Nagi Gebraeel, Kamran Paynabar, and A.P. Sakis Meliopoulos are with Georgia Institute of Technology, Atlanta, GA, USA.}}

\maketitle

\begin{abstract}
Complex interconnections between information technology and digital control systems have significantly increased cybersecurity vulnerabilities in smart grids. Cyberattacks involving data integrity can be very disruptive because of their potential to compromise physical control by manipulating measurement data. This is especially true in large and complex electric networks that often rely on traditional intrusion detection systems focused on monitoring network traffic. In this paper, we develop an online detection algorithm to detect and localize \textit{covert attacks} on smart grids. Using a network system model, we develop a theoretical framework by characterizing a covert attack on a generator bus in the network as sparse features in the state-estimation residuals. We leverage such sparsity via a regularized linear regression method to detect and localize covert attacks based on the regression coefficients. We conduct a comprehensive numerical study on both linear and nonlinear system models to validate our proposed method. The results show that our method outperforms conventional methods in both detection delay and localization accuracy.




\end{abstract}
\begin{IEEEkeywords}
Cybersecurity, ISO, State Estimation, Detection, Localization
\end{IEEEkeywords}
\ifCLASSOPTIONpeerreview
\begin{center} \bfseries EDICS Category: 3-BBND \end{center}
\fi

\IEEEpeerreviewmaketitle
\section{Introduction}

The growing digitization of smart grids and the infusion of IoT technologies has  exposed numerous cybersecurity vulnerabilities \cite{kushner2013real, case2016analysis}. Cybersecurity of smart grids is a topic that has been studied extensively. Examples of common areas of interested revolve around data confidentiality (eavesdropping, phishing, spoofing) and availability (flooding, denial of service, and distributed DoS) \cite{gunduz2018analysis}. A sizeable research effort is focused on cyberattacks that target data integrity in smart grid applications.  Data integrity cyberattacks refer to the manipulation of sensor measurements (namely, false data injections \cite{liu2011false}) and control actions such as in the case of replay \cite{delgado2015smart} and covert attacks\cite{smith2011decoupled} where often the attacker's intent is to alter normal system operations and cause physical damages. Data integrity cyberattacks can be especially disruptive to grid operations; consider for example the 2015 Ukrainian blackout \cite{case2016analysis}. Recent research has shown that data integrity attacks can successfully bypass conventional detection schemes such as the bad-data-detection \cite{liu2011false,mo2014detecting}. Consequently, this paper focuses on developing a cyberattack detection scheme aimed at the detection of such cyberattacks in smart grid applications.

Aside from detection, cyberattack localization is a major challenge in the smart grid. Due to the sheer scale coupled with complex and dynamic interactions between cyber and physical components of the grid, a cyberattack on one part in the network can propagate very rapidly. Hypothesis testing has traditionally been the de facto approach for detecting and identifying the locations of sensors with anomalous data signatures \cite{van1984hypothesis, caro2011multiple, asada2005identifying}. As the size of the network grows so does the number of sensors and the frequency of hypothesis tests required to identify the anomalous sensors. This can create significant statistical and computational challenges. Statistically, it is hard to make inferences (interpret the test statistics) on a large number of hypothesis tests, especially when they are dependent. The false alarm rate also increases and correction methods can be conservative. Computationally, with the number of affected sensors unknown, the number of hypothesis tests grows exponentially with the number of sensors. Thus, with a large number of sensors in the smart grid, hypothesis testing becomes intractable. Graph-based approaches were also proposed for localization. They often require hierarchical partition of the network and are often computationally expensive \cite{nudell2015real,he2011dependency}. These challenges provide an opportunity to develop an integrated detection and localization methodology that is statistically interpretable and computationally efficient. Specifically, we extract the data feature that contains information about both the abnormality caused by a covert attack and the location of the attack. The abnormality is identified by the magnitude, and the location is identified by the sparsity of the extracted feature.

\subsection{Related Work}
Models for detecting data integrity cyberattacks can be classified into two groups.
The first group uses network traffic from cyber communications to detect the attacks on the smart grid \cite{zhang2011distributed, liu2015abnormal,xu2017achieving}. These methods are based on detecting abnormalities in the network traffic data and are often similar in their operation to detecting DoS attacks.

The second group couples model-based detection with sensor data. A mathematical model is used to represent the normal behavior of the system. Attacks are detected using discrepancies between model prediction and actual system observations \cite{tan2017modeling,sridhar2014model, huang2018online,li2014quickest,zhao2018generalized}. Most of these methods are based on new designs of the detector, i.e., redefining the test statistic for detection \cite{tan2017modeling,sridhar2014model,li2014quickest,zhao2018generalized}. For example, in \cite{li2014quickest}, a CUSUM statistic is used to capture the cumulative error, and in \cite{zhao2018generalized}, the test statistic is extracted from the residuals of a robust state estimator. Another commonly used model-based approach is using authenticating data signatures to periodically verify the system's state \cite{huang2018online} Covert attacks where an attacker has sufficient knowledge of the system can still be missed using these approaches.

Literature related to localizing cyberattacks in smart grid applications is very limited. This is especially the case in data integrity attacks. One of the main challenges in localizing cyberattacks in smart grids is due to its complex physical interactions. An attack on one node can quickly propagate through the network making it difficult to associate the anomaly generated by the attack and its origin in the network. In \cite{nudell2015real}, a graphical model is used to locate attacks which are modeled as disturbances. However, it is not clear how this method can be extended to attacks that simultaneously manipulate the sensor data. In \cite{vukovic2013detection}, the attack localization is coupled with distributed state estimation, where each region shares its belief of the attack localization. In this paper, we use a centralized state estimation configuration to facilitate attack localization. The centralized approach utilizes the information from the neighborhood regions that can be affected by the attacked region to locate the cyberattacks more accurately.

\subsection{Contributions}

To the best of our knowledge, this is the first work that focuses on detecting covert attacks in smart grids. The paper focuses on power systems consisting of an Independent System Operator (ISO) and multiple Regional Control Centers (RCCs). We develop a method to detect and localize a covert attack on an RCC in (near) real-time.  This is accomplished by analyzing residuals from the ISO state estimation, in real-time. We demonstrate the effectiveness of our proposed methodology through a simulation study on an IEEE 14-bus system.

Our main contributions are summarized as follows: 1. We develop a generalized framework to model covert attacks on a regional control center. 2. We build an online covert attack detection mechanism by investigating and formally modeling the characteristics of the residuals from various sensor measurements under normal operations and under covert attacks. 3. We derive the impact of a covert attack on the neighboring regions of the targeted RCC.  This serves as the basis of our online attack localization scheme to identify which RCC is under attack. 4. Specifically, we leverage the unique sparse structure of the system residuals to locate the covert attack at the level of the individual generator bus.  This is achieved by utilizing Spare Group Lasso to enable efficient feature extraction.  The rest of the paper is organized as follows: In Section 2, we introduce the problem setup, including the system model, the attack model, as well as the Sparse Group Lasso (SGL) problem. In Section 3, we develop our methodology and propose two detection algorithms for linear and nonlinear systems, respectively. In Section 4, we conduct a numerical study and present the results. Finally, Section 5 concludes the paper.

    \begin{figure*}[h!]
    \centering
    \includegraphics[scale=0.4]{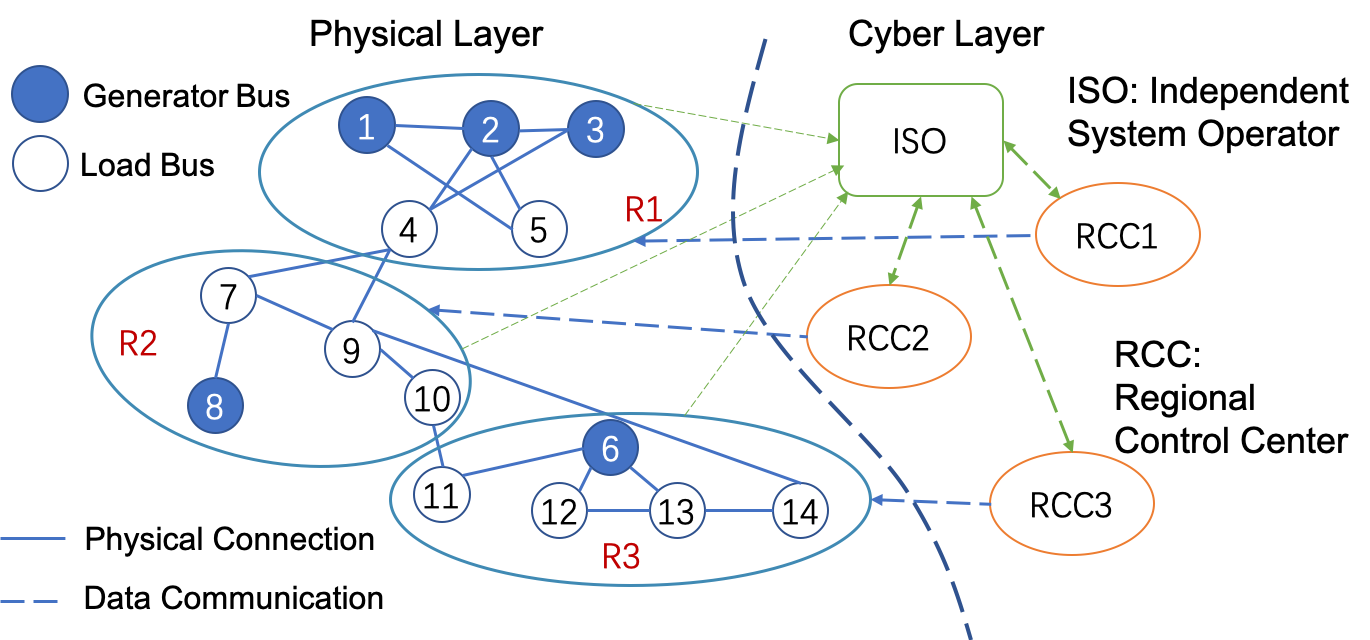}
    \caption{The Graphical Model of the Smart Grid (IEEE 14 bus)}
    \label{fig:network}
    \end{figure*}


\section{Problem Setup}
\subsection{System Model}
We consider an $N$-bus power transmission system comprised of power plants and substations that are grouped in $K$ different regions (an example of $N=14$ and $K=3$ is shown in Figure \ref{fig:network}). An ISO acts as a centralized coordinator which manages and controls the electric transmission of the power network. The power generation plants are operating under the control of regional control centers (RCCs). The global system state $\x \in \R^n$ ($n=2N-1$ under the $N$-bus setting) is defined as $\x^T=[x_1^T,...,x_N^T]$, where $x_i$ represents the state of bus $i$. In most cases, $x_i$ is defined as the voltage and phase angle for bus $i$. Without loss of generality, the phase angle of the reference bus is set equal to 0. Hence, $x_1\in\R$ when bus 1 is the reference bus, and $x_i\in\R^2$ for $i=2,...,N$. We assume there are $m$ ($m>>n$) sensors in the global system that guarantee the observability of the system. We denote the vector of measurements as $\z\in \R^m$, which may include the line power flows, the bus voltages and/or currents, loads on all the load buses, and the generation powers of all the generator buses. The measurement function is represented by a (nonlinear) measurement function

\begin{equation}
    \z=h(\x) \label{eq:nonlinear}
\end{equation}
the state estimation is given by solving the following optimization problem:
\begin{equation}
\min_\x (\z-h(\x))^T\Sigma^{-1}(\z-h(\x))\label{eq:lsq}
\end{equation}
where $\Sigma$ is the diagonal matrix of the sensor measurement precisions.

For a nonlinear model, the above problem is solved using Newton's method, and the state estimate $\hat{\x}$ is calculated iteratively as follows:
\begin{equation}
    \hat{\x}^{\nu+1}=\hat{\x}^{\nu}-(J^T\Sigma^{-1}J)^{-1}J^T\Sigma^{-1}(\z-h(\hat{\x}^{\nu}))\label{eq:newton}
\end{equation}
where $J$ is the Jacobian matrix of $h(\cdot)$, and $\nu$ is the iteration.

However, if the system operates around a state $\x_0$, the model can be properly linearized around $\x_0$ \cite{nudell2015real}. The state $\x_0$ can be obtained from a recent state estimation, which remains valid for multiple observations. The generalized linearization has the following form:
\begin{equation}
    \z=H\x,\label{eq:linear}
\end{equation}
where $\z$ and $\x$ are redefined as their linearization around $\z_0$ and $\x_0$, and $H$ is the measurement matrix derived from $\x_0$. In this case, the above state estimation problem is solved as a linear regression problem. That is, 
\begin{equation}
    \hat{\x}=(H^TH)^{-1}H^T\z.\label{eq:regression}
\end{equation}
This approach is commonly used in the literature. For example, a state-space model is used in \cite{huang2018online}, and a linear regression model is used in \cite{yu2017patopa}. 


\subsection{Covert Attack}\label{covert}
A covert attack is a cyberattack that maliciously manipulates system controls covertly by manipulating the sensor measurements. The mechanism of a covert attack on a steady-state linear system was first proposed by  \cite{smith2011decoupled}. Sensor data and control actions were manipulated by injecting two bias terms, which were assumed to be linearly dependent. In \cite{li2020degradation}, the authors proposed the covert attack against a linear dynamic system. The proposed attacks were proved to be undetectable when all the sensors in the system are vulnerable to manipulation. 

In contrast, this paper tries to introduce a covert attack on a nonlinear system. We provide a more generalized definition of a covert attack that is applicable to both linear and nonlinear systems. For both types of systems, we assume that an attacker has full knowledge of the system.  With this knowledge and access to controllers, the attacker can manipulate control actions arbitrarily. In addition, with access to the sensors, the attacker can simultaneously manipulate sensor data to disguise the control actions. We also assume that an attacker can only access one power plant at a time. In this work, we consider the following covert attack vector described below:

\vspace{0.05in}

\noindent 1. An attacker reads and manipulates the control of a  generator, $i$,  such that the state (power and/or voltage) of generator $i$ is altered. That is,
\begin{gather}
    x_i^a=x_i+\beta_i, \label{eq:attack1}
\end{gather}
where $x_i$ is the original state of generator bus $i$, $\beta_i$ is the shift of the state caused by a covert attack, and $x_i^a$ is the state of bus $i$ under attack.

\vspace{0.05in}

\noindent 2. The attacker simulates the expected sensor measurements corresponding to the ``normal'' state of the generator using knowledge acquired of the system model; 
\begin{gather}
    \z=h(x) \label{eq:attack2}
\end{gather}
\vspace{0.05in}
\noindent 3. The attacker manipulates the corresponding sensor measurements of generator, $i$, using simulated sensor measurements as follows, 
\begin{equation}
    \z_k^a=\z_k \quad \forall k\in M_i\label{eq:attack3}
\end{equation}
where $M_i$ is the set of sensors connected to generator $i$, including the ones that are measuring the state of bus $i$ and those that are in the neighborhood of bus $i$.

If an attacker was to orchestrate such an attack vector, it is highly unlikely that the attack would be detected by existing detection models. Note that all the sensors in $M_i$ that would have otherwise registered an abnormality in the state $x_i$ have been manipulated by the attacker, i.e., the readings indicate normal system behavior. The covert attack has proven to be undetectable for linear systems when the attacker has access to all the sensors and full knowledge of the system dynamics \cite{smith2011decoupled}. In \cite{van2015sequential}, the authors designed a detection method for scenarios where the attacker has limited access to the sensors, which is similar to our assumption here. However, the limited access implies that a subset of the sensors is always protected from manipulation. 

\begin{figure}[htbp]
     \centering
     \includegraphics[scale=0.43]{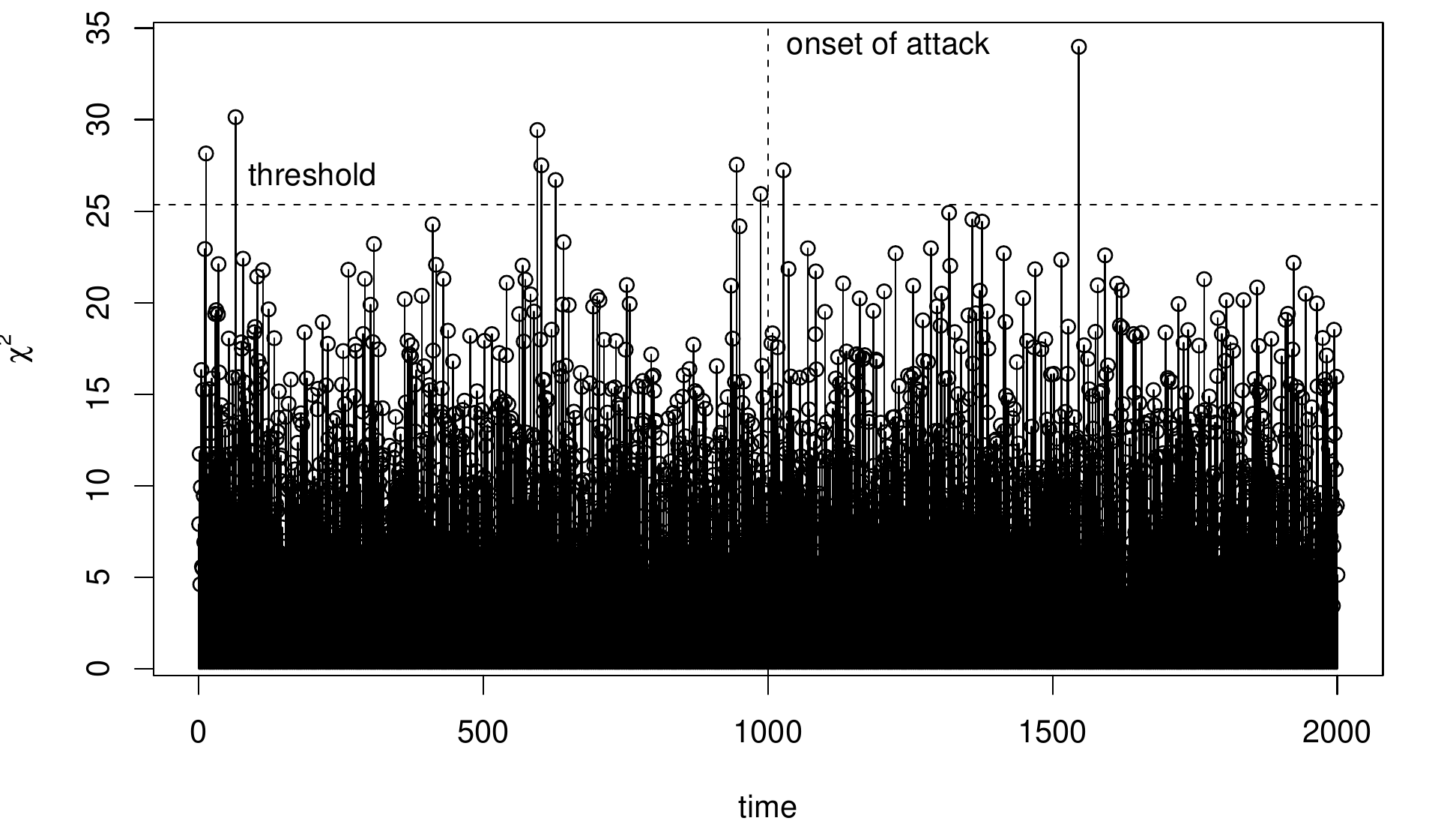}
     \caption{The $\chi^2$ detector fails to detect the covert attack}
     \label{fig:tsq}
     \vspace{-0.1in}
    \end{figure}

In contrast, this paper assumes that none of the sensors are immune to manipulation. Instead, we assume the attacker could only access sensors related to the targeted generator bus. We believe that this is a reasonable assumption since it is highly unlikely that an attacker would access all sensors in a large system simultaneously. By manipulating only the related sensors, the attacker is able to bypass traditional bad data detection schemes, such as $\chi^2$ detection. For example, in Figure \ref{fig:tsq} we plot the $\chi^2$ statistic of a system that is experiencing a covert attack since time $t=1000$. As shown in the figure, there is no significant change before and after the onset of the attack.

\section{Methodology}
In this section, we develop an online approach to detect and localize covert attacks in a smart grid. Our problem setting considers a smart grid comprised of multiple regions. Each region consists of generator buses that can experience a covert attack. For illustrative purposes, Figure \ref{fig:network} shows a smart grid comprised of three regions. We assume that an ISO collects all the sensor data from all the regions and that the topology of the network is known. \remove{ where our algorithm can be used to estimate the state of the grid. Our algorithm is designed to be implemented at the ISO, where all the sensor measurements are used, and the complete network topology is known.} Our detection and localization algorithm is coupled with the smart grid state estimation at the ISO level. 

In this paper, we consider a scenario where only one generator bus experiences a covert attack (note that our model can be extended to multiple busses). If the true state of the system is known, then residuals derived from sensor observations can be utilized to detect a covert attack. \remove{sensor data anomalies caused a covert attack. Here residuals refer to the differences between actual sensor observations and expected sensor measurements estimated using knowledge of the true state.} In this case, we expect that only the subset of sensors connected to the generator bus will display relatively large residuals. This will present significant sparsity in the residuals. Sparse Group Lasso (SGL) can be used to extract relevant (sparse) features that can detect and localize the covert attack. However, the true state of the system is typically unknown. Consequently, sparse features are used to correct the state estimation by coupling the observation model in Eq.(\ref{eq:nonlinear}) as a constraint to the SGL problem. This will be formalized later in Eq.(\ref{eq:sgl}). In the following subsections, we discuss the development of our methodology on a linear approximation of the system. Next, we relax the linear assumption and extend the methodology to the nonlinear setting.

\remove{We first introduce the formulation of the Sparse Group Lasso (SGL). Then, we show how to formulate the detection and localization problem as a constrained SGL problem based on a simplified linear model, in order to build the theoretical basis and develop the detection algorithm. Finally, we extend the methodology to the nonlinear system setting.}

\subsection{Sparse Group Lasso}
The formulation of sparse group Lasso was proposed in \cite{friedman2010note} as an advancement of the group Lasso problem that selects the group(s), among $L$ groups of predictors $X_1,...,X_L$, that explain the variation in the data $\y$
\begin{gather}
    \min(||\y-\sum_{l=1}^LX_l\beta_l||_2^2+\lambda\sum_{l=1}^L\sqrt{p_l}||\beta_l||_2),
\end{gather}
where $p_l$ is the group size, and $||\cdot||_2$ is the $L_2$ (Euclidean) norm. The penalty term $\lambda\sum_{l=1}^L\sqrt{p_l}||\beta_l||_2$ yields sparsity at the group level.
The sparse group Lasso considers within group sparsity in addition to the group level sparsity, which solves the following optimization problem:
\begin{gather}
    \min(||\y-\sum_{l=1}^LX_l\beta_l||_2^2+\lambda_1||\beta||_1+\lambda_2\sum_{l=1}^L||\beta_l||_2),
\end{gather}
where the $L_1$ penalty term $\lambda_1||\beta||_1$ yields the element-wise (within-group) penalty.
The SGL problem can be solved by block coordinate descent. The algorithm is given in \cite{friedman2010note}.

\subsection{Linear System}
We first hypothesize a simplified linearized model of the system, where the system operates around some state $\x_0$, and the measurement function is linearized in the form of 
\begin{equation}
    \z=h(\x)=H_0\x,\label{eq:ls1}
\end{equation}
where $H_0=h(\x_0)$ is a known constant, which is derived from the steady operating point $\x_0$. As given in Section \ref{covert},
\begin{equation}
    \z'=H(\x+\beta)=\z+H\beta,\label{eq:ls2}
\end{equation}
where $\z'$ is the measurements under covert attack \textit{before the attacker manipulates the sensor data}, and  $\beta$ is sparse, such that $\beta_i\neq 0$ and $\beta_j= 0$ for all $j\neq i$, because the attack only changes the state of generator $i$.
The measurements ($\z^a$) \textit{after the attacker manipulates the sensor data}, according to (\ref{eq:attack3}), is
\begin{gather}
    \z^a_j=\z_j, \quad\forall j\in M_i\label{eq:ls3}\\
    \z^a_j=\z'_j, \quad\forall j\in M^C_i\label{eq:ls4}
\end{gather}
where $M_i$ represents the set of sensors in the neighborhood of plant $i$, and $M_i^c$ is the complement of set $M_i$. From (\ref{eq:ls2}-\ref{eq:ls4}), we have
\begin{equation}
    \z^a=\z+B_i\beta_i,
\end{equation}
where $B_i\in\R^{m\times 2}$ is defined as follows:
$$B_i[M_i^C,\cdot]=H[M_i^C,S_i],$$
$$B_i[M_i,\cdot]=0,$$
with $B[S,\cdot]$ representing the rows in set $S$ of matrix $B$, and $B[\cdot,S]$ representing the columns in set $S$ of matrix $B$. In the above equations, $S_i$ denotes the set of elements corresponding to the state of bus $i$, $x_i$, in the state vector $\x$. In general, the matrix $B_i$ shows the relation between the state of bus $i$ and the measurements of the sensors that are not directly connected to bus $i$.

When the system is complex, each state is correlated with (different) multiple sensors. Notice that $B_i$ is obtained by taking some columns of $H$ and setting a subset of elements to zero. This transformation is nonlinear since it is element-wise and sparse, meaning there is a very high chance that $B_i$ is linearly independent of the column space of the matrix $H$. Therefore, when we know that bus $i$ is under attack, we could estimate $\beta$ in the following way:

1. Project the measurement onto the column space of matrix $H$ to estimate the state $\hat{\x}$ (i.e., solve the linear regression problem using (\ref{eq:regression}).

2. Project the residuals $\r=\z-H\hat{\x}$ onto the column space of matrix $B_i$, and the solution is $\hat{\beta}_i$, which can be expressed as:
$$\hat{\beta}_i=(B_i^TB_i)^{-1}B_i^T\r.$$

Note that in step 1, there might be the inaccuracy of state estimation because of leverage points. Therefore, the two steps should be done iteratively by removing the explained residuals ($B_i\hat{\beta}_i$) from $\z$ and re-estimating $\hat{\x}$. That is, in the second iteration, we first update the measurement as its correction by subtracting the explained residuals, i.e., $\z^c=\z-B_i\hat{\beta}_i$, and then estimate $\hat{\x}$. The residuals $\r$ in step 2 are still calculated based on the original measurement of $\z$. After we get the projection of the new residuals, we correct the measurement again using $\z^c\leftarrow\z^c-B_i\hat{\beta}_i$. These two steps are reiterated until the convergence of the estimated states.

The above solution is only valid when it is known that bus $i$ is under attack. When this is unknown, we need to locate the attack. Note that in the above solution, $B_i$ can be treated as the basis for bus $i$. Therefore, one can find all the basis for all the buses, and the problem can be formulated as finding the basis among $B_i$ for all $i$ that best explains the residuals $r$. Since it is likely that a subset of the states of bus $i$ is altered (e.g. when there are multiple generators in one power station, the attacker might only attack a subset of the generators, or the attacker only changes the power without changing the voltage), $\beta_i$ would be sparse. Therefore, if we divide the elements into groups according to the states of each node $i$, the estimated $\beta$ should be: 1) between-group sparse, meaning there should be only one basis $B_i$ that properly explains the residuals and 2) within-group sparse, meaning it is very likely that only a subset of the elements in $x_i$ is altered, which also means only a subset of the columns in basis $B_i$ is important in explaining the residuals variation.

The above problem can then be formulated as a Sparse Group Lasso (SGL) with linear constraint in the following form:
\begin{gather}
    \min_{\beta_i,...,\beta_L,\hat{\x}} ||\r-\sum_{i=1}^L B_i\beta_i||^2_2+\lambda_1 ||\beta||_1+\lambda_2\sum_{i=1}^L ||\hat{\beta}_i||_2\label{eq:sgl}\\
    \text{such that }
    H\hat{\x}+\r=\z,
\end{gather}
where $\lambda_1 ||\beta||_1$ is the $L_1$ penalty term that encourages within-group sparsity, and the $L_2$ penalty term $\lambda_2\sum_{i=1}^L ||\beta_i||_2$ encourages the between-group sparsity. Under our assumption that only one region is under attack, there is only one of all $||\beta_i||$'s that is significantly greater than $0$.
   \begin{algorithm}[h!]
   \label{alg:1}
     \caption{SGL-based attack detection and localization for linear system}
     \begin{algorithmic}[1]
     \renewcommand{\algorithmicrequire}{\textbf{Input: $tol$, $\z$, $H$, $\{M_1,...,M_K\}$, $\{S_1,...,S_N\}$, $\lambda$}}
     \REQUIRE
     \STATE $alarm=\textbf{0}$, $converge=\textbf{0}$;
     \FOR {$t = 1,2,...$}
        \STATE $\z=\z_{new}\leftarrow\z(t)$;
        \STATE $\hat{\x}_{old}=\x_0$;
        \WHILE {! $converge$}
          \STATE $\hat{\x}=(H^TH)^{-1}H^T\z$;
          \STATE $\r=\z-H\hat{\x}$;
          \STATE Solve (\ref{eq:sgl});
          \STATE $\z_{new}\leftarrow \z-\sum_{i=1}^L B_i\hat{\beta}_i$;
          \IF {($||\hat{\x}_{new}-\hat{\x}_{old}||<tol$)}
          \STATE $converge = \textbf{1}$;
          \ENDIF
          \STATE $\hat{\x}_{old}\leftarrow\hat{\x}_{new}$;
         \ENDWHILE
     \IF {($\max{||\hat{\beta}_i||_1} > \lambda$)}
      \STATE $alarm = \textbf{1}$;
      \STATE $location = \text{argmax}||\hat{\beta}_i||_1$
      \STATE \textbf{break};
      \ENDIF
     \ENDFOR
     \STATE \textbf{Return} $alarm$, $location$
     \end{algorithmic}
     \end{algorithm}
Since we assume that the basis $B_i$ does not lie in the column space of $H$, the above optimization problem can be solved iteratively by first solving the linear regression problem and then the SGL without the constraint, which is demonstrated in Algorithm 1. The solution $\hat{\beta}=[\hat{\beta}_1^T,...,\hat{\beta}_K^T]$ is an estimation of the attack vector $\beta$. For generator $i$ under attack, $||\hat{\beta}_i||>0$, and for generator $j$ not under attack, we expect to get $\hat{\beta}_j\approx0$. Within the detected generator, the none-zero elements would correspond to the altered state variables. When there is no attack, $||\hat{\beta}_i||$ would be close to 0 for all $i$. The online detection mechanism is built based on the maximum magnitude of L1 norm $||\hat{\beta}_i||_1$: the alarm is set when $\max||\hat{\beta}_i||_1$ is greater than the pre-specified threshold $\lambda$, where $\lambda$ could be selected based on the empirical distribution of $\max||\hat{\beta}_i||_1$ under normal condition. Specifically, $\lambda$ could be selected as the ($1-\alpha$) quantile of the empirical distribution of $\max||\hat{\beta}_i||_1$, where $\alpha$ is the desired Type-I error rate (usually set to 0.005). For generality, the $\max||\hat{\beta}_i||_1$ value is normalized based on the historical mean of $\max||\hat{\beta}_i||_1$ for the in-control data.

\subsection{Nonlinear System}
We now extend the formulation in the previous subsection to the nonlinear system setting as given by (\ref{eq:nonlinear}). In this case, the Jacobian matrix $H(x)$ is no longer a constant, but a function of $x$. Therefore, we need to re-approximate $H$ and the basis $B_i$ at every iteration, based on the new state estimation $\hat{x}$.

The detection and localization problem could be extended to a nonlinear system setting as a sparse group Lasso (SGL) problem with nonlinear constraints:
$$\min_{\beta_i,...,\beta_L,\hat{\x}} ||\r-\sum_{i=1}^L B_i\beta_i||^2_2+\lambda_1\sum_{i=1}^L ||\hat{\beta}_i||_1+\lambda_2\sum_{i=1}^L ||\hat{\beta}_i||_2$$
such that
$$h(\hat{\x})+\r=\z$$
$$B_i[M_i^C,\cdot]=H(\hat{\x})[M_i^C,S_i]$$
$$B_i[M_i,\cdot]=0$$
Note that the above optimization problem has no closed form solution or iterative algorithm with guaranteed convergence. However, we could linearize the constraint as follows. In practice, if the attacker alters the state of the generator by a large magnitude in a short time, it might directly shut down the generator, and the attack could be easily exposed. Therefore, we reasonably assume $\beta$ is not too large such that 
$$h(\x_a)\approx h(\x)+H(\x)\beta.$$
Similar to the linear case, $\beta$ is sparse with
$\beta_k=0 \quad \forall k\notin N_i$
As mentioned in Section \ref{covert}, the observed measurement $\z^a$ is an element-wise combination of $h(\x_a)$ and $h(\x)$. i.e.,
\begin{gather}
    \z'=h(\x_a)\approx h(\x)+H(\x)\beta,\\
    z^a_j=z_j', \quad \forall j\in M_i^C\\
    z^a_j=z_j,\quad \forall j\in M_i
\end{gather}
Since all the sensors that are directly related to the attacked node are covered by the normal measurements, the state estimation should be close to $x$. i.e., $\hat{\x}\approx \x$. Therefore, if generator $i$ is attacked, the residual $\r$ should satisfy:
$$r_j=H(\hat{x})[j,\cdot]\beta \quad \forall j\in M^C_i$$
which is equivalent to $$r=B_i\beta_i$$

Similar to the linear case, when region $i$ is under attack, the corresponding $\hat{\beta}_i$ should be large, otherwise we expect $\hat{\beta}_i\approx 0$.

The above optimization problem is solved iteratively by solving the state estimation and the SGL. At each iteration, we first solve the SE problem using Newton's method and get the residual $r$. Then, we solve the SGL using block coordinate descent and get estimates, $\hat{\beta}_i$, $i=1,...,L$. In the next iteration, the state estimation is solved by correcting $z$ using $\hat{\beta}_i$, $i=1,..,L$; i.e., $z'=z-\sum_{i=1}^L B_i\hat{\beta}_i$. At each time step, this procedure is iterated until convergence. We demonstrate the above procedure in Algorithm 2.

   \begin{algorithm}[htbp]
   \label{alg:2}
     \caption{SGL-based attack detection and localization for nonlinear system}
     \begin{algorithmic}[1]
     \renewcommand{\algorithmicrequire}{\textbf{Input: $tol$, $\z$, $H$, $\{M_1,...,M_K\}$, $\{S_1,...,S_N\}$, $\lambda$}}
     \REQUIRE
     \STATE $alarm=\textbf{0}$, $converge=\textbf{0}$;
     \FOR {$t = 1,2,...$}
        \STATE $\z=\z_{new}\leftarrow\z(t)$;
         \STATE $\hat{\x}_{old}=\x_0$;
         \WHILE {! $converge$}
          \STATE Solve $\min_\x (\z-h(\x))^T\Sigma^{-1}(\z-h(\x))$ using Newton's method;
          \STATE $\r=\z-H\hat{\x}$;
          \STATE Solve (\ref{eq:sgl});
          \STATE $\z_{new}\leftarrow \z-\sum_{i=1}^L B_i\hat{\beta}_i$;
          \IF {($||\hat{\x}_{new}-\hat{\x}_{old}||<tol$)}
          \STATE $converge = \textbf{1}$;
          \ENDIF
          \STATE $\hat{\x}_{old}\leftarrow\hat{\x}_{new}$;
         \ENDWHILE
         \IF {($\max{||\hat{\beta}_i||_1} > \lambda$)}
          \STATE $alarm = \textbf{1}$;
          \STATE $location = \text{argmax}||\hat{\beta}_i||_1$
          \STATE \textbf{break};
          \ENDIF
     \ENDFOR
     \STATE \textbf{Return} $alarm$, $location$
     \end{algorithmic}
     \end{algorithm}

\section{Numerical Results}
We validate the proposed detection algorithm on both linear and nonlinear systems. For the linear system setting, we model the complete system with a 20-variable linear time-invariant state-space model composed of 4 regions. For the nonlinear system setting, we use the IEEE 14 bus model and decompose it into three regions as shown in Figure \ref{fig:network}.

\subsection{Linear System}
For simulation on the linear system, we use the following discrete-time state-space model to represent the system operations:
\begin{gather}
    \x(t+1)=A\x(t)+B\u(t)+\e(t)\label{eq:ss1}\\
    \z(t)=H\x(t)\label{eq:ss2}
\end{gather}
where $\x$ and $\z$ are the system state and sensor measurement, respectively, as defined earlier, $\u$ is the control action that is calculated by the controller to keep the system state at target. (\ref{eq:ss1}) is the state-transition function, and (\ref{eq:ss2}) is the measurement function. We generate the state-transition matrix $A$ randomly as a positive-definite matrix with the largest eigenvalue less than 1 to guarantee stability, the control action $\u$ is calculated by a coupled linear-quadratic regulator\cite{LQG}. Note that $H$ is a sparse matrix where each state variable only affects a subset of the sensors. The non-zero elements are generated from a uniform distribution between 0 and 1. The sensors connected to generator $i$ are defined by the strong correlation between sensor $j$ and state $x_i$, where $||H[j,S_i]||_\infty>0.5$ means sensor $j$ is directly connected to generator $i$. The state-transition function is taken as a ``black box" which is assumed unknown, and the steady-state estimation is implemented only based on (\ref{eq:ss2}) using (\ref{eq:regression}). In this simulation, we have 20 state variables (i.e., $\x\in\R^{20\times1}$) and 30 sensors (i.e., $\z\in\R^{30\times1}$). We run the detection algorithm for $N=500$ replications to evaluate the detection delay and localization performance on average. In each replication, the attacked region $i_{\text{attack}}$ is chosen randomly, and the thresholds remain unchanged.

The attack detection performance of our proposed method is evaluated by the in-control and out-of-control average run length, i.e. $ALR_0$ and $ALR_1$. The average run length is defined as the average number of observations before an alarm is raised:
\begin{equation}
    ARL=\E[\min{\{t: \max{||\hat{\beta}_i(t)||_1} > \lambda\}}]\label{eq:arl}
\end{equation}

It has the following relation with the type-I and type-II error rates:
\begin{gather}
    ARL_0\approx\frac{1}{Pr(\text{type-I error})}=\frac{1}{Pr(\text{false positive})}\\
    ARL_1\approx\frac{1}{1-Pr(\text{type-II error})}=\frac{1}{1-Pr(\text{false negative})}
\end{gather}
We choose the threshold $\lambda$ offline based on the empirical distribution of $\max||\hat{\beta}_i||_1$, such that the Type-I error rate $\alpha=0.005$, so the expected in-control average run length is 200. The magnitude of attack is defined by the signal-to-noise ratio (SNR):
$$SNR=\sqrt{\beta_i^T\Sigma_i^{-1}\beta_i},$$
where $\Sigma_i$ is the covariance matrix of the state variables of generator, $x_i$.

The out-of-control $ARL_1$'s along with their standard deviations under different SNR's are shown in Table \ref{tab:arl}. For comparison, we use the traditional $\chi^2$ detector as a baseline, whose ARL is also given in the table.

Recall that in our proposed algorithm, the attack localization is identified as $\text{argmax}||\hat{\beta}_i||_1$. The attack localization performance of the proposed method is evaluated by the identification accuracy, precision, recall, and the $F$ score of the proposed attack localization approach, which are calculated using the following equations:
\begin{gather}
    \text{Accuracy}=Pr(\text{argmax}||\hat{\beta}_i||_1=i_{\text{attack}}),\label{eq:acc}\\
    \text{Precision}=\frac{TP}{TP+FP},\\
    \text{Recall}=\frac{TP}{TP+FN},\\
    F=2\cdot\frac{\text{Precision}\cdot\text{Recall}}{\text{Precision}+\text{Recall}},
\end{gather}
where $TP$, $TN$, $FP$, $FN$ are the number of true positives, true negatives, false positives, and false negatives, respectively. The precision values in Table \ref{tab:arl} is obtained by taking the average of the precision values over all three regions, and the same applies to the recall and $F$ score values.
The accuracy represents the probability that the algorithm correctly identifies the attacked region at the time of detection. Precision represents the proportion of correct alarms among all the alarms, recall represents the proportion of correct alarms among all the cases where region $i$ is indeed under attack, and $F$ score is the harmonic mean of precision and recall.

The accuracy of attack localization of the proposed method is compared with a modification of the hypothesis testing technique used in the literature \cite{van1984hypothesis}, where we test the group of sensors that are related to region $i$ all together. 
More specifically, for each region $i$, we remove the sensors in the set $M_i$ and re-estimate the state using the remaining sensor measurements. The new $\chi^2_i$ statistic is calculated accordingly. After we go through all the regions, the new $\chi^2_i$ statistics are compared, and the attacked region is identified as the region $i$ that minimizes $\chi^2_i$. This is because a low $\chi^2_i$ value means the removed sensors best explains the abnormality. The accuracy of the proposed method and the hypothesis testing method is given in Table \ref{tab:arl}.

The results in Table \ref{tab:arl} show that the proposed method has a higher detection power and a higher localization accuracy than the traditional $\chi^2$ detector. For example, when the SNR is 1, the localization accuracy of SGL is 69.8\%, which is greater than the accuracy of the $\chi^2$ detector, 48.8\%. When the SNR is 6, the accuracy of both the methods increase, where the $\chi^2$ detector reaches an 86.6\% accuracy, and SGL reaches a 99.4\% accuracy, which is also better than $\chi^2$. Table \ref{tab:arl} shows as SNR increases, the localization accuracy for both methods increases. However, the proposed method has higher accuracy than the hypothesis tests under all the tested SNR levels. More importantly, the proposed method reaches a reasonably high accuracy (more than 85\%) at a relatively low SNR level (SNR=2), while the hypothesis test reaches similar accuracy at a much higher SNR level (SNR=6). This means the proposed method is more sensitive to covert attacks.

\begin{table*}[htbp]
  \centering
  \caption{ARL and Accuracy under different levels of SNR}
    \begin{tabular}{c|ac|ac|ac|ac|ac}
    \toprule
          & \multicolumn{2}{c|}{ARL(Std.Dev.)} & \multicolumn{2}{c|}{Accuracy} & \multicolumn{2}{c|}{Precision} & \multicolumn{2}{c|}{Recall} & \multicolumn{2}{c}{F score} \\
    \hline
    SNR   &$\chi^2$ & SGL   & $\chi^2$ & SGL   & $\chi^2$ & SGL   & $\chi^2$ & SGL   & $\chi^2$ & SGL \tstrut\bstrut\\
    \hline
    0     & 203.46 (8.04) & 200.84  (9.96) & -     & -     & -     & -     & -     & -     & -     & - \\
    1     & 171.68 (7.26) & 153.34 (6.80) & 48.80\% & 69.80\% & 52.74\% & 72.63\% & 48.80\% & 69.80\% & 47.35\% & 69.61\% \\
    2     & 97.404 (4.48) &  82.78 (3.88) & 60.80\% & 85.80\% & 63.13\% & 86.77\% & 59.45\% & 85.37\% & 59.91\% & 85.92\% \\
    3     & 50.20 (2.22) & 41.47 (1.95) & 68.20\% & 92.60\% & 73.86\% & 92.79\% & 68.27\% & 92.60\% & 68.51\% & 92.69\% \\
    4     & 23.32 (1.00) & 16.18 (0.66) & 72.40\% & 97.00\% & 75.97\% & 97.13\% & 71.37\% & 97.05\% & 72.18\% & 97.08\% \\
    5     & 15.54 (0.63) & 13.11 (0.51) & 80.40\% & 96.40\% & 82.90\% & 96.77\% & 80.91\% & 96.38\% & 81.28\% & 96.58\% \\
    6     & 12.13 (0.44) & 8.38 (0.27) & 86.60\% & 99.40\% & 87.99\% & 99.39\% & 86.36\% & 99.41\% & 86.91\% & 99.40\% \\
\bottomrule
    \end{tabular}%
  \label{tab:arl}%
\end{table*}%

\begin{figure}[ht!]
     \centering
     \begin{subfigure}[b]{0.48\textwidth}
         \centering
         \includegraphics[width=\textwidth, height=2.1in]{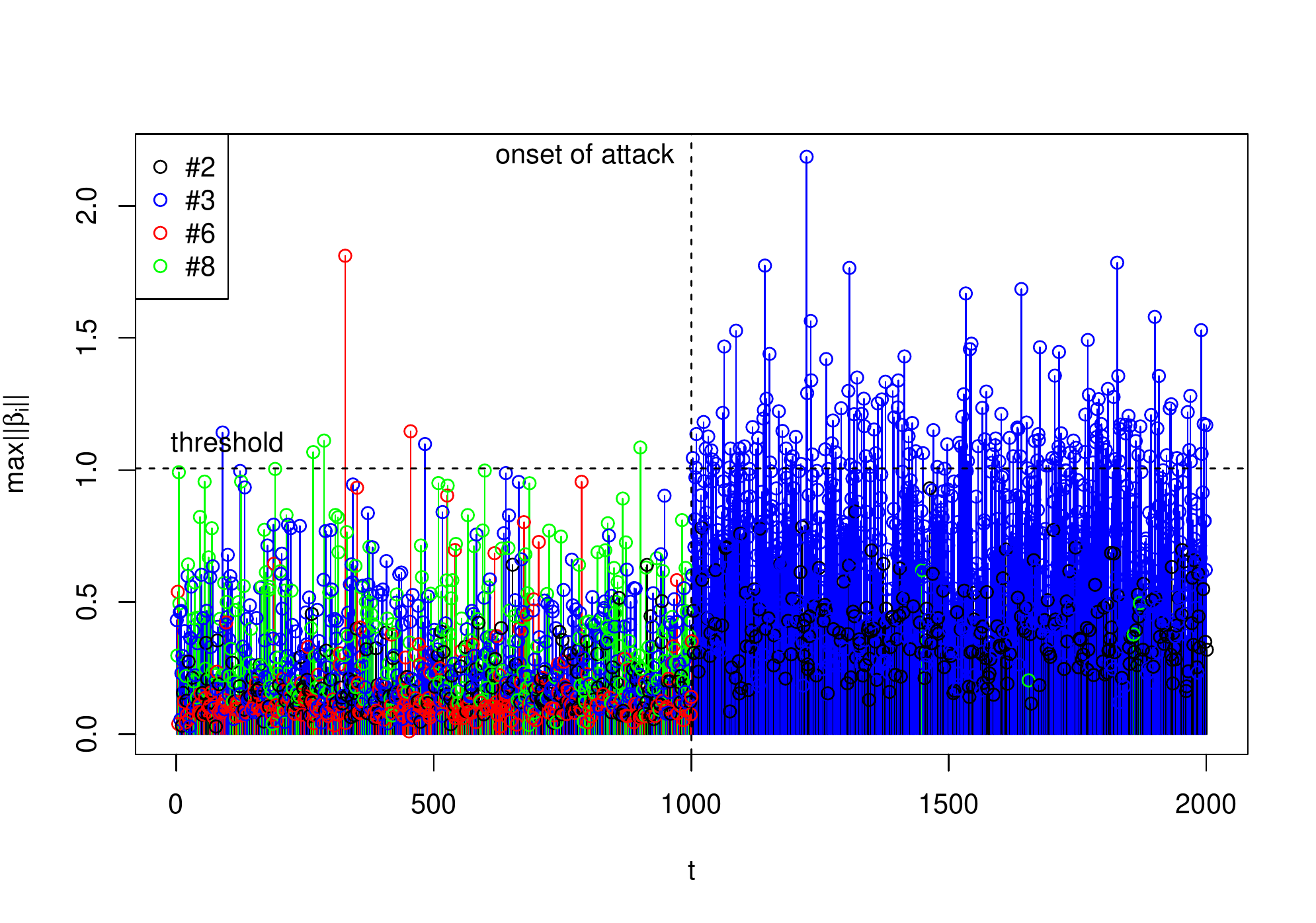}
         \caption{Detection of attack on \#3}
         \label{fig:result3}
     \end{subfigure}
     \hfill
     \begin{subfigure}[b]{0.48\textwidth}
         \centering
         \includegraphics[width=\textwidth, height=2.1in]{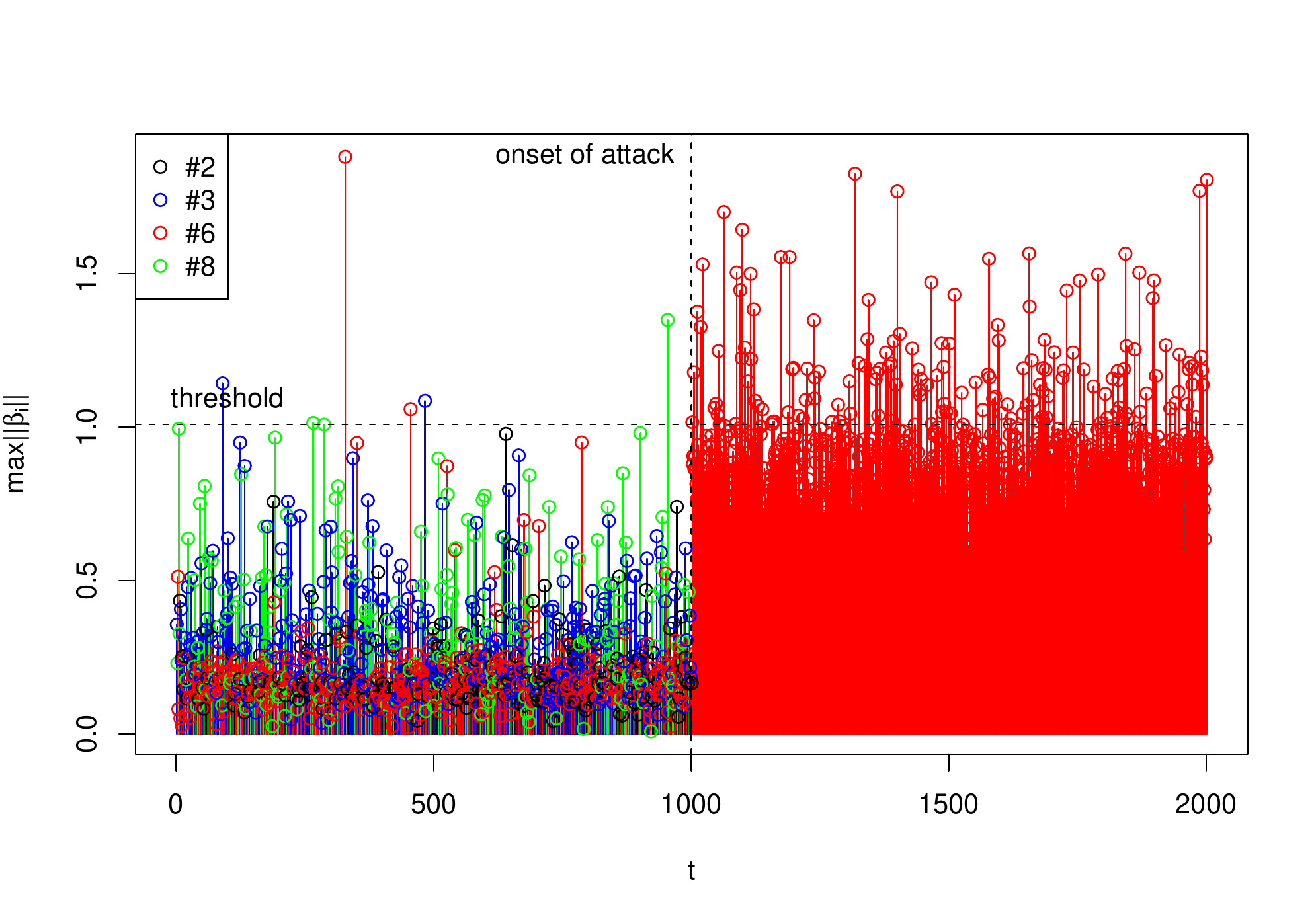}
         \caption{Detection of attack on \#6}
         \label{fig:result6}
     \end{subfigure}
        \caption{Simulation Results}
        \label{fig:examples}
    \end{figure}
    
    \begin{figure*}[htbp]
     \centering
     \includegraphics[width=0.6\textwidth,height=1.5in]{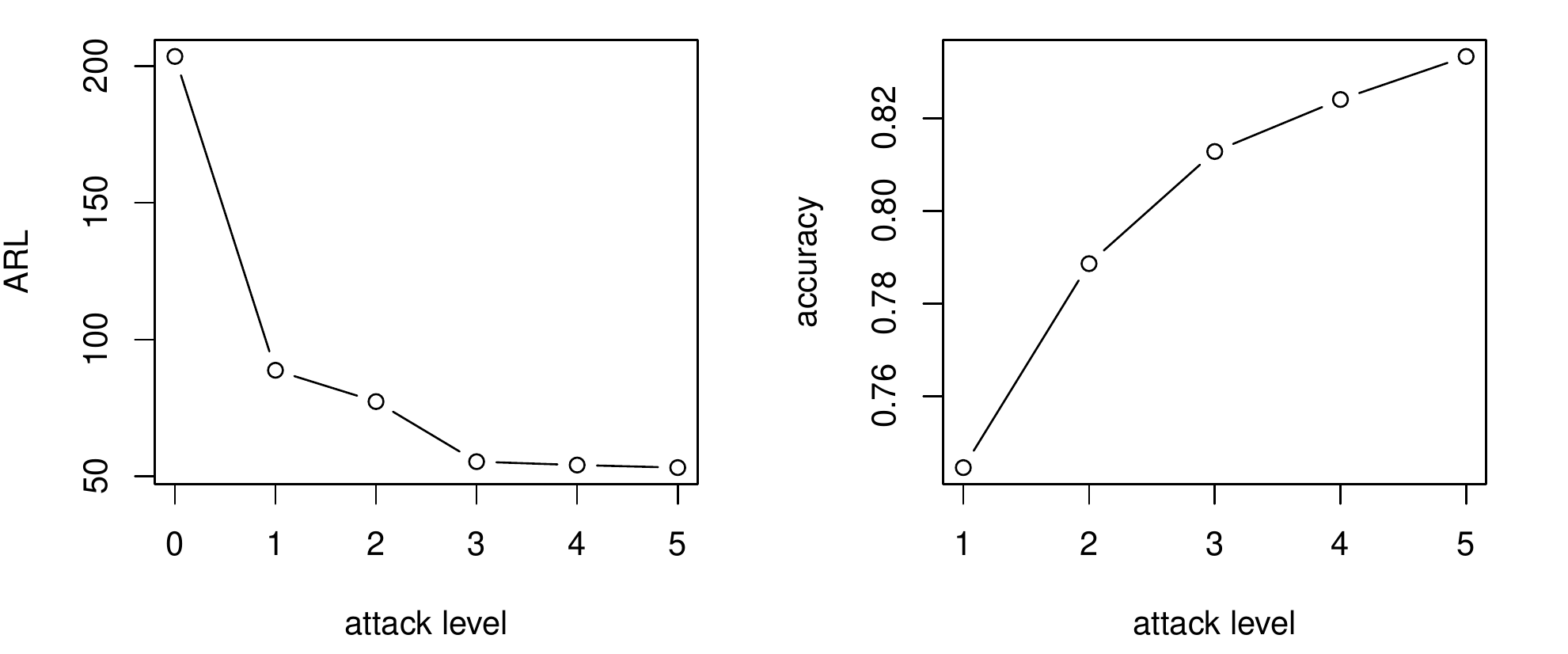}
     \caption{ARL and localization accuracy under 5 levels of attack}
     \label{fig:arl}
    \end{figure*}
\subsection{Nonlinear System}
To validate the performance of the method on nonlinear systems, we simulate the attack using the IEEE 14 bus model. The input to the simulation is the load profiles of the load buses, the generation plan of the generator buses, and the phase angles and voltages at each bus. The load profiles are generated from the real data from Pecan Street dataset. The generation plan is generated based on the load by solving the mixed-integer unit commitment problem \cite{carrion2006computationally}. 

We simulate the system for 2000 observations and monitor the $l_1$ norm of the $\hat{\beta}_i$ vectors. The threshold is selected based on the $0.995$ quantile of the monitoring statistic ($\max ||\hat{\beta}_i||$), such that the in-control average run length is around 200. There are 5 levels of attack, where level 1 to 5 represents decreasing the generation level by 20\% to 100\%. For each level of attack, we replicate the simulation 500 times, and in each replication, a covert attack on one of the generators randomly selected among buses \#2, \#3, \#6, and \#8. 

Two examples of attack on \#3 and \#6 are shown in Figure \ref{fig:examples}, where the attack occurs at $t=1000$. Before the onset of the attack, the magnitudes of $||\hat{\beta}_i||$ for all generator buses generally follow the same distribution, and false alarms are triggered with a low possibility. On the contrary, after the onset of the attack, the $||\hat{\beta}_i||$ values for the region under attack is much higher than the others, and the $\max ||\hat{\beta}_i||$ values are above the threshold such that alarms are frequently triggered.

We evaluate the performance of the method using average run length and the localization accuracy defined by Equations (\ref{eq:arl}) and (\ref{eq:acc}). We show the variation of the average run length and the localization accuracy under the 5 levels of attacks in Figure. \ref{fig:arl}. The results show that, as the attack level (severity of attack) increases, the detection delay decreases, and the localization accuracy increases. This means the proposed method has a higher detection power as well as a higher localization accuracy as the attack is more severe.

\section{Conclusion}
We proposed an online approach to detect and locate a type of data integrity attack called a covert attack. Our detection approach is based on the SGL formulation. We showed the theoretical foundation of applying SGL to a linear system setting and extend the method to a nonlinear system setting with relaxation. We conducted a simulation study to evaluate the performance of our proposed method by highlighting the average run length, which indicates the expected detection delay, as well as the localization identification accuracy. The results showed that as the severity (SNR) of attack increases, both the detection power and the localization accuracy of the proposed method increase. The results also showed that the proposed method is much more sensitive than $\chi^2$ tests. Furthermore, we implemented a case study on the IEEE 14-bus system as a representative of the more practical nonlinear systems. The results showed that the proposed method is applicable to nonlinear systems and able to reach shorter detection delay and higher localization accuracy as the attack severity increases. As future work, we will investigate the scalability and the robustness of the proposed method. We can also extend the method to detection, localization, and identification of other types of cyberattacks and faults in smart grids. Another direction is to incorporate the deep-learning based methods in order to reach a high accuracy and still preserving interpretability. Presently, our optimization algorithm is research grade. Our plan is to further develop the code into a commercial grade optimization algorithm that solves the SGL with linear and/or nonlinear constraints.

\bibliographystyle{plain}
\bibliography{ref}

\end{document}